# Sculpt, Deploy, Repeat: Fast Prototyping of Interactive Physical Objects

**Michael Jones**
BYU Computer Science
3328 TMCB, Provo, Utah
Mike.jones@byu.edu

**Kevin Seppi**
BYU Computer Science
3328 TMCB, Provo, Utah
k@byu.edu

**ABSTRACT**

We present a method for sculpting deployable prototypes of interactive physical objects—which we call "PhysiComps." Building a deployable PhysiComp that merges form and function typically involves a significant investment of time and skill in digital electronics, 3D modeling and mechanical design. We aim to help designers quickly create prototypes by removing technical barriers in that process. Other methods for constructing PhysiComp prototypes either lack fidelity in representing shape and function or are confined to use in the studio next to a workstation, camera or projector system. Software 3D CAD tools can be used to design the shape but do not provide immediate tactile feedback on fit and feel. In this work, sculpting around 3D printed replicas of electronics combines electronics and form in a fluid design environment. The sculptures are scanned, modified for assembly and then printed on a 3D printer. Using this process, functional prototypes can be created with about 4 hours of focused effort over a day and a half with most of that time spent waiting for the 3D printer. The process lends itself to concurrent exploration of several designs and to rapid iteration. This allows the design process to converge quickly to a PhysiComp that is comfortable and useful.

**ACM Classification Keywords**
H.5.2. Input devices and strategies

**INTRODUCTION**
The advent of inexpensive 3D printers and cheap microelectronics makes it possible to think about and experiment with new shapes for computing that are better adapted to our physical needs. The form factors of workstations, laptops, tablets and smart phones need no longer constrain the shape and size of a computing device. We can now think of interactive computing embedded in toys, tools and other everyday objects that fit into our lives more subtly and flexibly than was previously possible. We refer to such physical computation objects as PhysiComps.

We see the PhysiComp design space as quite large, ranging from traditional appliance or automobile control to measuring spoons that talk to digital cookbooks, cups that report the heat and volume of their contents, cooking spoons that measure temperature and stirring effort, and walking canes that report gait and falls for the elderly. The goal is to embed sensing, computation and communication into everyday objects in ways that make them more useful and comfortable.

We are intrigued by the challenge of integrating 3D physical shape into the process of designing interactive computation. One of the largest barriers we see is not cost or feasibility but rather the human effort required to produce a prototype of a PhysiComp. There are still too many skills required. Though the materials and manufacturing costs are now low, one still must have good skills in digital electronics, 3D modeling, 3D printing, signal processing and mechanical design.

When designing a PhysiComp, the affordances of the 3D shape are critical. It is not enough to make a device that functions, the device must both function and feel right. How it "fits in my hand," "sits comfortably on my arm," "slides in and out of my pocket," or "is comfortably reached on the steering wheel" are all critical questions when designing physical interactions. Traditionally the shape is created and evaluated by carving foam, wood or some other material with successive shapes being tested for fit and feel. Digital interaction with the object is virtually excluded until very late in the process because of the difficulties of rapidly integrating digital electronics, 3D shape and actuator mechanics.

Hudson and Mankoff [8] created a rapid prototyping technique based on simple materials such as cardboard, thumbtacks and tape. This allowed them to create physical prototypes that were fully interactive in a matter of minutes. This is the kind of rapid turn around that is necessary in PhysiComp design. The resulting prototypes, however, did not have a realistic feel, nor were they sufficiently robust for actual deployment. The Calder toolkit [9] is closer in that it is based on foam to give it a more realistic shape and to preserve rapid design time, but the controls are still crude. The prototype is tethered and not deployable in a realistic way. We merge realistic shape and realistic interaction earlier in the process and do so in a device that is not tethered to a workstation.

Physical shape and its relationship to interaction techniques and usage scenarios requires many iterations to create an effective result. Our goal is to reduce the time to deploy a functional prototype of a new PhysiComp from weeks and months down to days. Fast iterations are the norm in screen-based interactive software. We have a rich body of tools and techniques for rapidly creating, evaluating and modifying visual interaction designs. Such tools and techniques do not exist in the 3D shape domain.

We want a prototype that can be actually deployed into use so that its usability can be understood. This creates a fundamentally different design cycle where shape and interaction are integrated from the beginning.

As a guiding example in pursuing this goal we consider the contributions of reusable widget sets such as the widgets in X windows or the Macintosh toolkit. Each widget encapsulated a fully implemented interaction in a form that was easily reused and adapted for a variety of purposes. The widget set also provided simple tools for integrating individual widgets into the whole of the user interface.

Our approach to simplifying and speeding up the PhysiComp design cycle includes sculpting, scanning and printing the prototype. The prototype is then outfitted with electronics and deployed for use. Sculpture as a medium for exploring user experiences in 3D as presented here is analogous to sketching as a medium for exploring user experiences in 2D [4].

The prototype can be used in the intended context and the process repeated for another design iteration. Repeated designs can also happen concurrently when multiple designs are sculpted and deployed together.

Sculpting, scanning and printing provide tangibility, personalization and composition that are more difficult to achieve with software CAD tools. Sculpting naturally provides tactile feedback that is not provided by a 3D CAD tool. Personalizing a shape to a person or fitting a PhysiComp, including its embedded electronics, into an existing object is as easy as pressing clay into shape.

## Components of a PhysiComp

Figure 1 shows the basic components of a PhysiComp. Sensors collect information about user activities. There are a large variety of possible sensors. In this paper we experiment with a variety of electronic sensors including an optical flow sensor, a camera, accelerometer, and a gyro as starting points. In addition, there is a processor that collects and classifies sensor data in order recognize user activity and input commands. In many of our designs the sensor data is transferred out of the object for processing elsewhere. However, there are other designs where activity is communicated directly to the physical world via actuators such as lights and servos. Another challenge that cannot be ignored when using digital electronics is power.

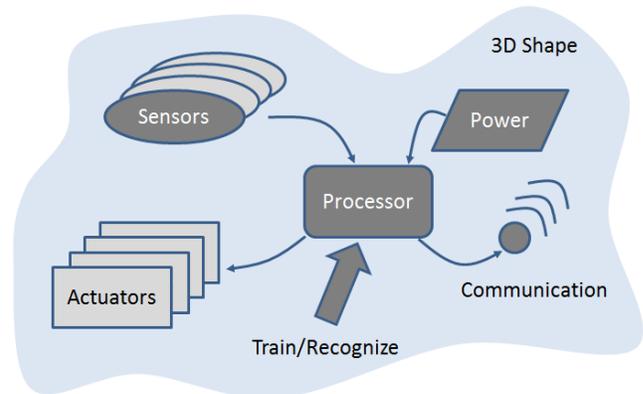

**Figure 1 – Components of a PhysiComp.**

## Embedding interaction into shape

All of these components must be encapsulated in a physical 3D shape. Adapting the device shape to the electronics form factor and adapting the electronics form factor to the device shape is one of the challenges of creating such devices. We also find that creating shapes which fit naturally to the user's experience in a specific sutation is a non-trivial part of the design process. Too many personal devices and appliances have shapes that are not as humane as they could be. We want to empower a design esthetic that subordinates computation and electronics to shape and experience. In that sense our aims match the ideals in Vertegaal and Holman's [7,13] organic user interface design but we place more emphasis on merging physical form with interaction rather than painting displays on existing objects.

An important part of the prototype design process that is frequently overlooked is assembly and disassembly. It is not enough to have the right shape that contains the electronics, but that shape must be easy to assemble. Creating PhysiComps presently is a "death by a thousand cuts." There are many little things for which there are well known solutions but which must all be dealt with in order to create a robust prototype. It is not that any one of them is complex or hard in isolation but that there are so many of them. One of the goals of our project is to bring design/construction decisions down to the absolute minimum so that the focus is on the design of shape and experience.

To achieve this, the mounting fixtures for the electronics must be embedded in the shape design. There must be a clear means for disassembly and reassembly. For assembly there should be the necessary clips, holes, pins or other devices to make that work. All of these things must be taken into account while designing, manufacturing and assembling the shape. These trivial but necessary details must fit naturally and simply into the design process.

## PRIOR WORK

### Prototyping Functionality and Form Together

There is a long history of development of prototype tools for physical interaction. Some work, like the .NET Gadgeteer [14], have focused on the barrier of digital electronics. The Gadgeteer provides a board with a universal pluggable bus. This eliminates hardware development and greatly simplifies software development, but the form factor is so large that it is unusable for many PhysiComp devices.

Other work focused on craft-based materials as a platform for exploring physical devices. The "untoolkit" uses paper and fabric along with pens with conductive ink to design new devices [10]. This approach simplifies the creation of digital electronics but does not teach us much about prototyping deployable objects. Lilypad [3] takes a similar approach using fabric and conductive thread.

In Hudson's BOXES project [8], cardboard and thumbtacks allow rapid prototyping of button placement on a physical shape. BOXES allows for very rapid ideation and fits well in that niche. Our aim is to support a wider range of electronic systems and to allow a more accurate representation of shape while maintaining BOXES' focus on the merging form and function early in a fluid process.

The Switcheroo [2] and Calder [9] projects use interactive devices that can be pinned to foam models. This allows for rapid modification of the shape and an instant test for how it feels in the hand or next to the body. The foam provides the shape and devices provide the interaction. This is closer to the kind of prototype that we want, except that there is no allowance for embedding electronics inside the shape and devices stuck on the outside made not accurately model the final form and interaction.

More recently, Sauron [12] provided a way to prototype form and functionality using more complex 3D shapes. They develop their 3D shapes in SolidWorks. The results are less humane than we would like because CAD tools provide no tactice feedback like clay and foam do. Sauron allows interactive objects to be placed on the surface and uses a video tracking system with a camera inside the device to detect interaction events like button presses or knob turns. The system is tethered to a PC by a cable that transmits video from inside the device.

Our work differs from Sauron in that we use sculpture rather than a CAD tool to generate the device shape and we allow for a wide range of electronics inside the device. We expect that the Sauron widgets and sensing system could be used in our process by sculpting around 3D printed representations of Sauron's widgets and internal electronics.

### Adding Functionality to Existing 3D Forms

A number of projects have used cameras and projectors to create the illusion of physical interaction or to augment physical objects as part of a prototyping process. We focus on adding functionality to new objects but share a common goal to explore physical interactive objects. A key difference is that our objects are not confined to a studio with projectors or cameras but can be used anywhere.

Instant user interfaces [5] uses depth cameras to sense the position and orientation of physical objects. From this information interactive behaviors can be recognized. This pushes interaction out into the user's physical space.

WorldKit [16] does not use physical objects for interaction but allows the user to "paint" interaction onto physical surfaces. Projectors provide the visual instantiation of the interactive object and cameras sense the interaction. The interaction is restricted to the surface of existing objects. SLAP widgets [15] also interact on surfaces, using a camera for the interaction. Transparent physical objects are used to interact, while projectors provide the visual feedback.

Paper Windows [6] use Vicon cameras to track paper-like objects and then project images onto those objects. The resulting interaction feels like one is manipulating interactive paper. This idea is extended in DisplayObjects[1]. Instead of paper, foam objects with actual three-dimensional feel are instrumented with Vicon tracking balls and projector can then project the user interface onto the surface of the object.

All of these systems rely upon a closed interactive world. There must be cameras for the sensing. Occlusion from the camera eliminates the interaction. There must also be one or more projectors that inject the visuals into the physical environment. These too have occlusion problems. This work fills a niche of "what would it be like if we could build it?" In thinking ahead, these systems can prototype what cannot yet be and provide a sense of future behavior. However, these cannot be realistically deployed for people to actually use in their lives.

The Midas toolkit [11] is much closer to a deployable prototype but does not strongly merge form and functionality in the process. Midas provides a tool for drawing the shapes of various touch-based interactors. These shapes are then modified into the appropriate electronic objects using copper tape and a vinyl cutter. The result is a set of custom touch surfaces that are fully interactive and can be stuck onto the surface of physical objects. The interaction is restricted to touches on a 2D surface but the result is very close to interactive deployment and feels very much like the end result. The object can be deployed into the real world. For the purposes of our work, Midas offers a kind of interaction that is easy to use with a sculpted prototype by simply sticking copper tape on the sculpture. We used the Midas approach to make the touch-based hiking headlamp prototype described later.

Holman and Vertegaal [7] argue persuasively that organic, humane shapes are key to the future of interactive devices. People's hands and bodies must fit with the interactive objects that we create. Vertegaal [13] later emphasized the importance of humane shape and specific function in "organic user interfaces." Our work shares the foundational idea that 3D objects with a specific function must fit the user in a specific environment. But we consider a more general class of electronic systems and we merge form and function into a single process rather than covering existing objects with high resolution displays.

**PROTOTYPE FABRICATION**
Our fabrication process is comprised of the following steps:

1) 3D print a blank copy of the embedded circuit joined with the mounting shapes, power supply, and the exposed interaction surfaces, if any. These can be printed in advance for impromptu use.

2) Embed the 3D blank in clay or foam and then sculpt a 3D shape that is appropriate for the desired application.

3) Scan the sculpted shape using a 3D scanner.

4) Convert the scanned shape to create a hollow shell separated into at least two pieces and add necessary mounting shapes that will correctly place the electronics and the power supply.

5) 3D print the shell

6) Add the electronics to the printed shell along with the power supply and assemble the pieces into our desired device. Train a gesture recognizer if needed.

7) Try the device with users

8) Repeat (1-8) until satisfied.

For the purposes of this project we are ignoring the programming of our devices. Once inputs are generated by the user they simply become events for code that can be implemented in traditional ways. We use wireless communication protocols to communicate with a laptop that holds the software functionality. This allows us to focus on shape and its relationship to user controls. Admittedly there is some level of user testing that is not possible with this approach and will depend on fully embedding the functionality into a processor in the device.

### Running Example
In the following discussion we will use an optical flow microcamera from an optical mouse as a running example but mention generalizations of the process where appropriate.

This optical flow sensor creates a very rich design space when used, not against a desktop but rather turned over and stroked with a finger, or other body part to generate recognizable signals. By focusing on a sensor that is not often used we can explore the design process and the relationship between shape and interaction in a simplified setting.

We used boards from a Logitech M187 Optical Wireless mouse, which includes on optical flow sensor and a wireless communication package. The boards are about 38 mm by 51 mm with the sensor on one end. We modified the board by replacing the AAA battery mount with a mount for a smaller watch battery positioned on the center of the board.

The microcamera in the mouse senses motion 2 mm from the lens opening across a diameter of 14 mm. However, the microcamera *accurately* senses motion across a diameter of 13 mm. The microcamera and board were mounted at a 10 degree angle relative to the viewing plane of the camera, on a 2mm sheet of Plexiglas.

The physical sensing limitations of the microcamera impose constraints on the physical placement of the input window on the PhysiComp surface and the microcamera on the circuit board. These constraints must be satisfied or the device will not capture gestures.

**CREATING PHYSICAL SHAPE**
We designed our first PhysiComps using 3D CAD tools such as Rhino and SolidWorks to design prototypes. This proved unsatisfactory for several reasons. First, it is critical that a shape physically fit with human use. If the device is to be held in your hand it must fit there naturally and comfortably. This natural fit is very difficult to achieve without holding the shape during the process.

Second, it is tedious to create human form fitting shapes using solid geometry primitives and operations such as extrude, revolve and loft. Nuts, bolts and machine parts are made to fit these kinds of shapes, but human bodies are not. Figure 2 shows two sculptures molded to fit a hand. Such shapes would be difficult to create as quickly using lofted or extruded surfaces.

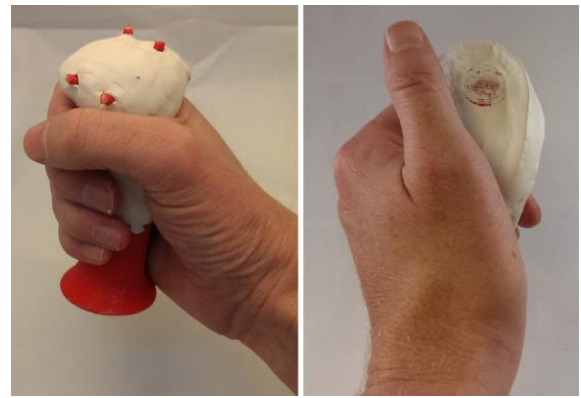

**Figure 2 – Sculpting enables comfortable placement of the thumb relative to exposed interactive surfaces.**

We settled on sculpting as a means for creating PhysiComp shapes. We tried 5 kinds of clay and found that Polyform Sculpy Polymer clay provided a nice balance between pliability and plasticity and could be hardened. Plastilina clay was not pliable enough and does not harden which makes scanning difficult.

While modeling in clay allowed us to create comfortable shapes quickly we initially had great difficulty in fitting the electronics and mounting bracket inside of our newly designed shapes. This is further complicated by the fact that the exact placement of some electronics relative to the PhysiComp surface can be critical. For example, the optical mouse microcamera had to be placed at a specific position and orientation relative to the interaction window. In other cases, such as PhysiComps that only sense motion, placement relative to the surface is less critical.

Precisely preserving relative placement of optical flow camera, the interaction window and the finger during early design phases could distract from free form exploration of the device shape.

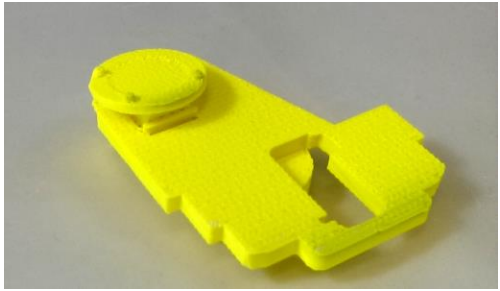

**Figure 3 – A 3D printed blank representing circuitry.**

We address this by sculpting around a 3D printed representation of electronics components and interaction elements. We sculpted around a 3D printed "blank" which was an analog representation of a correctly configured circuit board and interaction window. The printed blank for the optical flow camera is shown in Figure 3. The instructions to the designer are that the interaction portion (which is the round circle) must be exposed and flush with the surface and other parts of the blank may be exposed or protrude from the shape but that the blank must not be carved, sanded or cut off. Exposed interaction elements gave the designers a clear idea of how interaction placement integrates with the shape.

The printed blanks include mounting brackets and are expanded by 3mm in all directions to allow for the thickness of the walls of the PhysiComp shell. Expanding the blank by 3mm allows designers to safely leave the blank exposed while leaving room for the shell wall. If the blanks were not expanded to allow for the shell wall the designer might sculpt down to the blank only to discover that 3mm of material must be added back on to the design to allow for the wall. Figure 4 shows where a designer has just exposed part of the circuit blank and stopped.

In the case of the mouse camera optical flow sensor we did not expand the geometry that represents the interaction window because we want to place the actual window precisely on the surface with correct size and shape.

For the mouse optical flow sensor, the blank also included three bumps on the round interaction window (as seen in the upper left of Figure 3). These bumps are used after scanning to recover the alignment and position of the circuit board within the sculpted shape. Adding the three bumps eliminated the need to recover the circuit and mounting bracket orientation within the sculpture. Other interaction elements that did not have the radial symmetry of a perfect circle did not require additional markers to recover orientation.

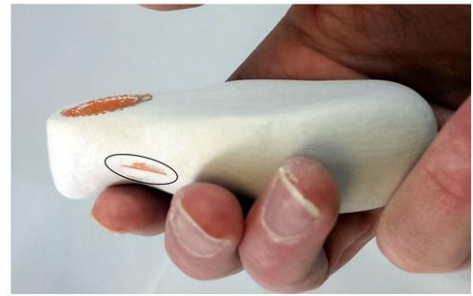

**Figure 4 – Sculpting down to the blank.**

## CREATING DIGITAL SHAPE

The starting point for the digital shape is created by scanning the sculpture. We created 3D scan of sculptured devices using a NextEngine 3D scanner. We found that the most important factor in creating good scans was to position the sculpture so that it nearly fills the camera frame. We also drew small x-shaped targets on the surface to assist later in aligning fragments of the scan.

Ideally, scanning produces a closed triangular mesh of the shape. In practice, cleanup using a tool such as Netfabb Professional was required to close the shape and repair the errors mesh representation (by eliminating exposed edges and inverted faces).

Meshes could be smoothed after scanning but smoothing clay before scanning made scan fragments easier to assemble, resulted in meshes with fewer errors and required fewer polygons. Even a few minutes spent smoothing the sculpture using a wet sponge or finger improved scanning.

Once the outer shape is in watertight digital form, it can be converted into an "assembly." The conversion into an assembly is done in a 3D CAD tool. Conversion involves the following tasks:

1. Shelling and splitting the scanned shape,
2. adding mounting brackets for the circuit and a hole for the interaction window, and
3. adding fasteners for securing the pieces together.

The exterior of the digital shape is an exact copy (up to the limits of the 3D scanner) of the sculpted shape

### Shelling and splitting

A hollow shell is created in one step by using the shell command in Netfabb Professional. Shelling commands in other tools such as SolidWorks or Rhino produce meshes with intersecting faces that cause problems for Boolean operations on meshes and for 3D printing. Most likely because the shelling algorithm in these tools produces an offset surface by moving vertices along vertex normals which can result in self-intersecting faces that need to be removed later. We found that shells with a thickness of 3 mm made durable prototypes while supporting mounting brackets and fasteners added later in the process.

The shell is split by defining a plane that intersects the object. In many cases the splitting plane is parallel to the interaction window but this is not essential.

**Adding the Mounting Bracket**

We created a mounting bracket model for each circuit that could be inserted into shapes by extruding the back face of the bracket to the inner surface of the shell. As mentioned earlier, for the gesture sensor built from a mouse microcamera optical flow sensor, it is challenging and important to match the alignment the bracket within the 3D model to the placement of the printed blank within the sculpture. We placed the bracket in the model by creating the plane defined by the three bumps on the interaction window. This plane and the bump positions define the location and orientation of the mounting bracket. A series of Boolean subtractions create the hole for the interaction window.

The mounting bracket is rotated 10 degrees along the longest dimension of the bracket relative to the face of the hole on the object surface and placed to align with the window location and orientation. Again, recovering the circuit orientation is a critical factor in this step and the three bumps printed onto the optical flow circuit blank simplify this process.

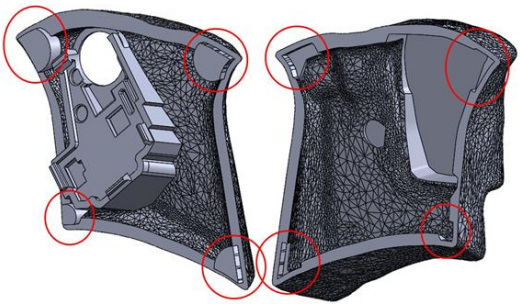

**Figure 5 – Fastener design.**

**Adding the Fasteners**

Next, geometry is added to fasten the printed pieces together. We experimented with thin pins (3mm diameter and 5 mm height) and holes. These pins and holes held the pieces together while allowing about a millimeter of lateral movement when closed but tended to break easily. We also used a tongue and groove closure laid out along the entire rim of each piece, but this did not securely join the pieces.

The fastener design shown in Figure 5 securely joins the printed pieces and is not easily broken. The fastener consists of a boss and a hole that are joined in a friction fit. Bosses are added to the piece containing the circuit board mounting bracket. Matching holes are added to the other piece. Bosses are created by drawing a closed curve that intersects the piece and lies on the plane used to split the shape into two pieces. The curve is then extruded 3 mm above (away from the piece) the cutting plane and down to the piece surface. Holes are created by rejoining the two pieces along the cut plane and performing a Boolean subtraction.

The process of shelling, splitting, placing the mounting bracket and the assembly bosses takes about 25 minutes for an experienced creator using our standard shape libraries. We automated this process for the mouse microcamera optical flow sensor.

**ASSEMBLING THE PROTOTYPE**

The digital shape is exported for printing, printed and assembled to create a functional prototype. For objects based on the mouse optical flow sensor, assembly includes putting together the printed pieces of the shell, a Plexiglas window and the circuit board. These pieces are shown in Figure 6. The Plexiglas window has a diameter of 20 mm and a 13 mm circle engraved onto the surface. The 13 mm circle provides tactile feedback that indicates the limit of the optical flow sensor. The Plexiglas window is glued to a 2mm rim around the hole in the printed shell.

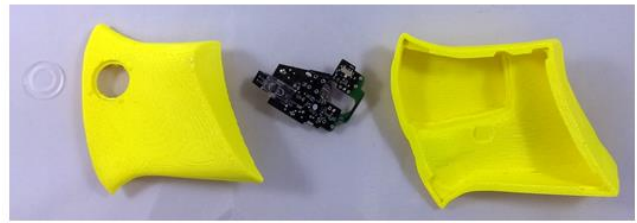

**Figure 6 – Parts for assembly.**

No additional effort is required in assembly to obtain the correct alignment of the interaction window and optical flow sensor. The mounting bracket and interaction window were placed in the digital design so snapping the circuit into the bracket will produce the correct placement.

Assembling PhysiComps based on other electronics was also simple. However cables and wires connecting different components for other electronics proved difficult to handle without adding representations of wires and cables to the printed blanks. Wire and cable representations modeled positioning constraints such as cable lengths, shapes and orientations.

We exported the digital shape as an STL file for printing. Pieces were printed on one of two different 3D printers. Low-resolution prints were created using a Stratasys Dimension Elite printer with a layer thickness of 0.254 mm. Higher resolution prints were fabricated on a Stratasys Objet30 printer with a layer thickness of 28 microns. Both printers use removable support material. Assembly and use of printed pieces was identical in both cases. However, pieces printed on the Objet30 felt more like the clay original.

## INTERACTION

The focus of our work is creating deployable prototypes of objects that support the intended user interaction. Like the task of physical design, our approach to prototyping interaction also seeks to lighten to load on the designer. We have focused on touch and gesture interaction. Touch and gesture have no moving parts would add a new set of mechanical design considerations for 3D printing but support a wide range of interactions.

We explored three modes of gesture interaction with the mouse microcamera optical flow sensor: one-handed, one-fingered and two-handed. Examples of each of these three modes are shown in Figure 7. One-handed gestures are created in hand-held devices such as a TV remote control or ski pole handle using the thumb. One-fingered gestures are created on a surface that is part of an object not held in the hand such as a blender or alarm clock. Two-handed gestures are created when both hands hold an object such as a steering wheel or bicycle handlebars.

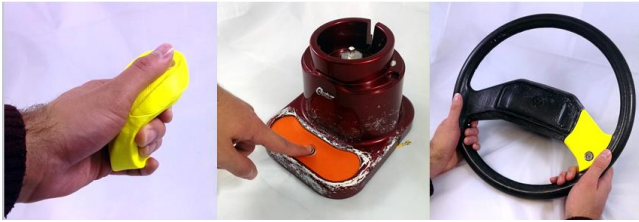

**Figure 7 – Three different gesture positions.**

In other prototypes we used a touch screen or a capacitive touch button based on copper tape. The touch screen was large and rectangular (7.11 cm diagonal) which lead to large and rectangular PhysiComps. Copper tape was considerably more flexible in size and shape but required a bit more assembly to connect a wire to the copper tape and then to the circuit board (a Red Bear Lab Blend Micro Arduino).

### Generating Gesture Input

Gesture input using a finger and a mouse microcamera optical flow sensor requires correctly positioning the finger relative to the camera. The finger must remain within the field of view and the depth of field. Our sensors had a depth of field between 1.4 and 2.1 mm and sensed motion most accurately 2 mm from the lens. To get this right requires interaction with a real shape.

We experimented with several interaction window designs for placing the finger correctly over the mouse microcamera. Several square and round holes of size 2 to 7 mm printed in the object surface did not support accurate gesture capture because it was difficult to keep the finger surface 2 mm from the sensor lens. When the finger is pressed against a hole, the skin protrudes into the hole at a distance that varies with pressure. It was difficult for users to simultaneously move their fingers while keeping pressure constant. Interaction windows covered with Plexiglas windows kept the finger surface positioned 2 mm from the lens even with variations in pressure. A window with diameter 14 mm supported a wide range of motion.

This is an example of one of a "thousand cuts" when trying to build prototypes. For our designers, these issues are handled automatically because these sensor geometry issues are built into the shape of the blank.

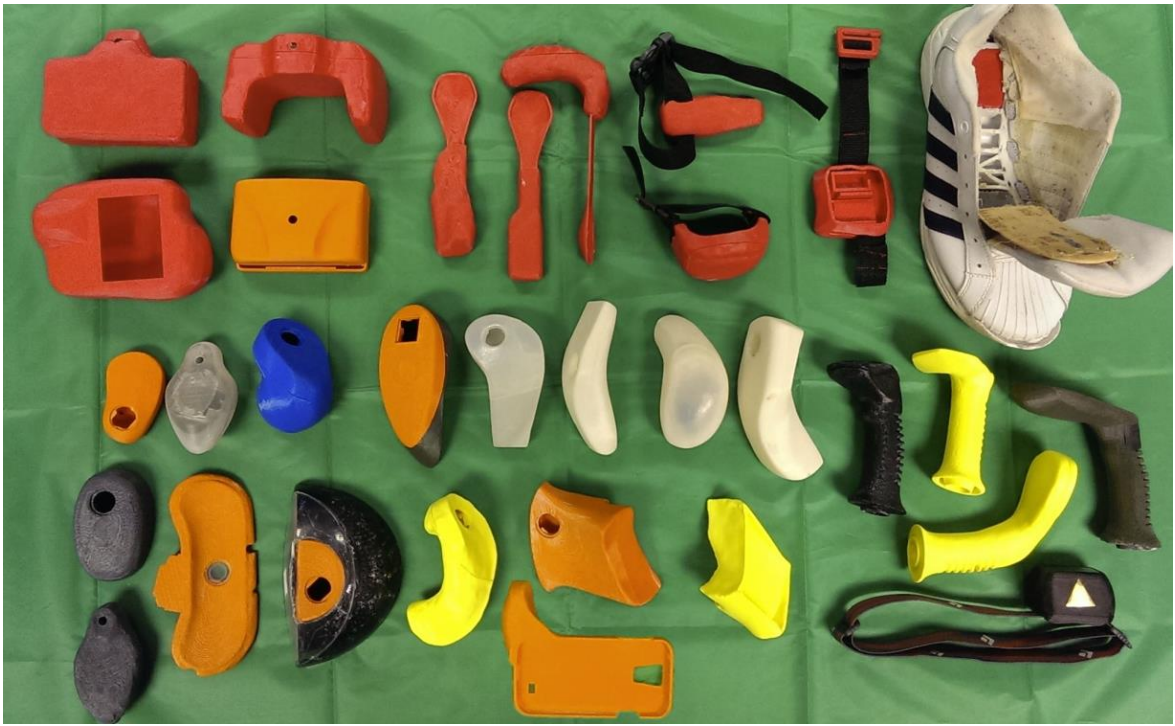

**Figure 8 – PhysiComp prototypes created using the sculpt, scan and print process.**

Classifying gestures is another barrier to creating a functional prototype. We simplified gesture recognition by training a neural network (with one hidden layer containing 20 nodes and trained with back propagation) to classify strokes. The neural network learned different classifiers for each of 5 different devices with 89 to 94 percent accuracy using a total of 120 samples generated by 4 different people in 15 minutes. Classifiers for one device did not transfer to other devices.

**DESIGN STUDIES**

We have conducted several design studies involving 9 different designers working with 9 different design scenarios involves 4 circuit platforms. We asked most designers to sculpt multiple designs for multiple scenarios. A total of 32 prototypes were sculpted and fabricated in these case studies and are shown in Figure 8.

**Raspberry Pi Camera**

The four prototypes in the top left of Figure 8 were sculpted around a Raspberry Pi computer with a camera and touch screen. The touch screen and computer board were large and dominated the shape of the design. Each shell contains a small round hole for the lens and as large rectangular hold for the touch screen. The small holes visible in three of the shells is the lens hole and the large rectangle in the other shell is for the touch screen.

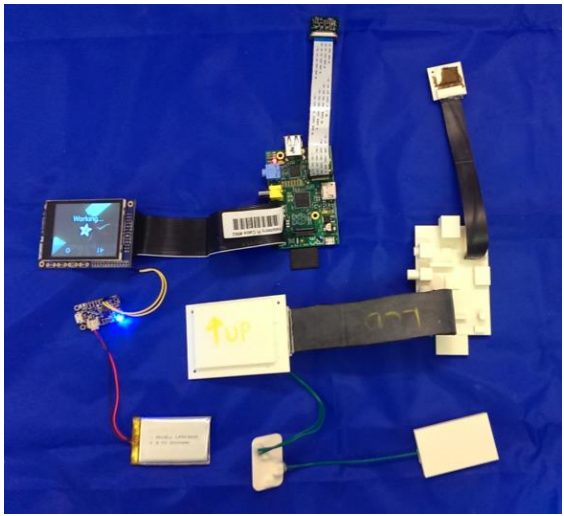

**Figure 9 – Camera based on the Raspberry Pi together with a 3D printed representation of the circuit.**

Designers explored unique camera possibilities in which the lens and screen were not aligned. They felt that this made it easier to view the image preview while taking a picture. Interestingly, these designers recalled early view camera designs in which the camera with held at waist level and a preview shown on ground glass.

The printed blank for use in sculpting is shown next to the actual electronics in Figure 9. We printed 3D models of the computer itself, a battery, the touch screen and the camera. We included extra geometry to represent pin headers and cable connectors. Flexible rubber from a bicycle inner tube represented flexible ribbon cables connecting the screen and camera to the computer. This mimicked the flexibility and the volume of the ribbon cables during sculpting.

**Data Logger**

Prototypes in the top right of Figure 8 were sculpted around a data logger. These include the three spoons in the center of the top row, the three bracelets with nylon straps and the red device in the white shoe. These do not have interactive surfaces and were intended to hold a sensor during a specific activity such as eating, walking or exercising.

Working with wet clay led to quickly constructed shapes that housed the sensor while allowing users to wear the sensor comfortably. Bracelet designs were quickly customized by simply pressing wet clay against the users' wrist. The design was comfortable for other users but made a perfect fit against the original user's wrist.

The shoe insert was also easy to sculpt even though the final geometry was complex. Figure 10 shows a closer view of the shoe insert along with the original sculpture. Sculpting the shell for the shoe insert was a matter of pressing wet clay into the shoe sole structure. The resulting shape fit perfectly and required little effort to construct.

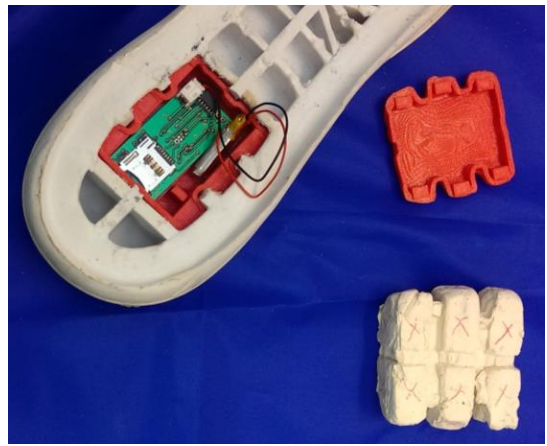

**Figure 10 – Prototyping a shell for a data logger in the sole of a shoe by pressing clay into the shoe itself.**

The printed blank for the sensor included the sensor itself and a battery connected by a 3D printed wire. The printed wire could be bent without breaking to allow for different placements of the battery relative to the sensor.

**Mouse Microcamera Optical Flow Sensor**

All other prototypes, except for the cell phone case in the bottom center and headlamp in the lower right corner, were sculpted around the mouse microcamera optical flow sensor as described in earlier sections of this paper. The middle row contains remote control designs on the left and ski pole handle designs on the right. The bottom row contains two remote controls on the far left and other controllers in the middle. Three of these are also shown in Figure 7.

For these prototypes, sculpting supported careful placement of the thumb directly over the sensor window. Designers could easily remove or add clay to get the thumb in the right place while creating a comfortable shape that also accommodated the circuit board.

The printed blank for the mouse optical flow sensor included the circuit and the interaction window. Because placement of the window relative to the camera was constrained to a single position and orientation the entire assembly was printed as a single object.

### Headlamp

The headlamp prototype is shown at the bottom right corner of Figure 11 and has a bright triangle on top. Both the headlamp and the cell phone case, which we discuss next, were single design case studies that met a specific purpose. The purpose of this design was to test a capacitive touch interface based on sticking copper foil to the prototype.

The copper foil acts as a button that turns the light on or off. Sculpting with clay then pressing adhesive copper foil onto the sculpture allowed the designer to explore different button shapes and placements. Using foil during sculpture, rather than drawing the button shape onto the sculpture ensures that the final shape can be cut from a sheet of foil.

The prototype was sculpted around the 3D printed representation of a circuit taken from a commercial off the shelf headlamp combined with a small Arduino board. The Arduino board monitors the capacitive touch button and switches power to the headlamp circuit. Because the circuit is not grounded, such as through a USB cable, two copper foil surfaces are needed with one wired to ground and the other to a general purpose IO pin. The user must touch both for a touch to be sensed.

### Cell Phone Case

The cell phone case is at the bottom center of Figure 8. This prototype simplifies using a cell phone to take a picture of a page from a book. The user can hold the phone in one hand and hold the book open with the other.

It was important and easy to place the user's thumb over the onscreen controls for the cellphone camera phone while maintaining a secure and comfortable grip on the phone holder. At one point the designer adjusted the grip handle angle by bending wet clay while holding the phone case.

Rather than print a 3D representation of the phone, we printed a phone case. The phone case acted as the mounting bracket geometry as used in the other design studies. We then scanned the first case with the sculpted handle to create a second case that worked as a prototype. The handle on the printed prototype was less likely to separate from the phone case than the clay handle on the plastic phone case.

## DISCUSSION OF DESIGN EXPERIENCES

We asked 6 of the designers to keep notes while designing some of the prototypes discussed above. Not all designers kept good notes for all designs but we did record the time spent in different phases of the process.

We found that both sculpting the PhysiComp and modifying the scanned shell in a CAD tool required the most time in focused attention. Printing required more time but did not require focused attention. We were comfortable with the time spent sculpting because that is the creative part of the process. Time spent using the CAD tools was frustrating for the designers and was viewed as a kind of overhead cost.

### Sculpting

Sculpting required between 10 and 165 minutes of focused effort. We were pleased that the sculpture process accommodated designers with varied backgrounds and always resulted in a functional prototype. Experienced designers (4 students in the final two years of an undergraduate industrial design program) used the most time and used that time to refine and perfect their shapes. Novice designers (2 students in an undergraduate computer science program) required the least time and left their shapes rough but functional. All designers were able to create shapes that felt right in the hand and functioned as expected.

The most time consuming sculpture was the steering wheel prototype which took 165 minutes. Most of that time was spent figuring out how to embed clay within the steering wheel cut out. The next most time consuming sculpture required 120 minutes to make the new blender interface. And most of that was spent embedding clay into the blender base. The shoe sensor, which also involved pressing clay into an existing object took only 20 minutes.

Designers who were also experienced CAD tool users (the industrial design students) noted pros and cons. Modifying a design in a CAD tool requires less effort than changing a clay sculpture. Changing a sculpture required moving actual physical material while changing a CAD design does not. But it was easier get the right physical fit and feel in clay than in a CAD tool. Scanning a sculpture results in a triangular mesh which is more difficult to work with in CAD than a parametric design created directly in a CAD tool.

### Scanning

Scanning required 5 to 30 minutes. Designers had little to say about the scanning process. Prior to the study we wrote instructions for the scanning process and designers followed these instructions with little interest or difficulty.

### Modifying the Scan

Modifying the shape in a CAD tool required between 20 and 210 minutes of effort and produced the most frustration. We used NetFabb to hollow the sculpture and Solidworks for adding mounting brackets and holes for interaction surfaces. Novice users expressed frustration in this phase. A common refrain was "I want place the mounting bracket right there, how do I make it go right there?" while pointing at the screen.

Experienced designers spent less time with the CAD tool, presumably because they had more experience. Most of their time was spent refining the placement of mounting brackets to fit in the shape while preserving the correct surface position and orientation of interaction elements.

Based on these experiences, we wrote a SolidWorks macro that automatically placed mounting brackets for the mouse optical flow sensor. The designer clicks three points on the surface of the interaction window and the macro completes the modifications included a placement optimization step. This macro reduced frustration and time for novice and expert users. Similar marking schemes and macros could be made for placing other electronics.

**Printing and Assembly**
Printing and assembly also required little of the designer's time and generated no commentary. Printing required several hours but required only a few minutes of focused attention.

**CONCLUSION AND FUTURE WORK**
The process allows a designer to create a functional PhysiComp prototype in about a day and a half with less than 4 hours of focused effort. The process supports fluid design and provides tactile feedback on both shape and interactive surface placement. Next steps include designing and allowing for widgets that include moving parts.